\documentclass[a4paper,conference,final]{IEEEtran}

\usepackage{amsmath}
\usepackage{amssymb}
\usepackage{balance}
\usepackage{cite}
\usepackage{color}
\usepackage{graphicx}
\usepackage{fancyhdr}
\usepackage{multirow}
\DeclareMathOperator*{\argmax}{arg\,max}

\IEEEoverridecommandlockouts

\graphicspath{{./}{figures/}}	

\begin{document}

\title{Large-Scale Location-Aware Services in Access: Hierarchical
  Building/Floor Classification and Location Estimation using Wi-Fi
  Fingerprinting Based on Deep Neural Networks}
  
\author{%
  \IEEEauthorblockN{Kyeong Soo Kim\IEEEauthorrefmark{1}, Ruihao
    Wang\IEEEauthorrefmark{1}, Zhenghang Zhong\IEEEauthorrefmark{1}, Zikun
    Tan\IEEEauthorrefmark{1}, Haowei Song\IEEEauthorrefmark{2}, Jaehoon
    Cha\IEEEauthorrefmark{1}, and Sanghyuk Lee\IEEEauthorrefmark{1}}%
  \IEEEauthorblockA{%
    \IEEEauthorrefmark{1}%
    Department of Electrical and Electronic Engineering\\%
    Xi'an Jiaotong-Liverpool University\\%
    Suzhou, 215123, P. R. China\\%
  }%
  \IEEEauthorblockA{%
    \IEEEauthorrefmark{2}%
    Department of Computer Science and Software Engineering\\%
    Xi'an Jiaotong-Liverpool University\\%
    Suzhou, 215123, P. R. China\\%
  }%
}%

\maketitle


\begin{abstract}
  One of key technologies for future large-scale location-aware services in
  access is a \textit{scalable indoor localization technique}. In this paper, we
  report preliminary results from our investigation on the use of deep neural
  networks (DNNs) for hierarchical building/floor classification and floor-level
  location estimation based on Wi-Fi fingerprinting, which we carried out as
  part of a feasibility study project on Xi'an Jiaotong-Liverpool University
  (XJTLU) Campus Information and Visitor Service System. To take into account
  the hierarchical nature of the building/floor classification problem, we
  propose a new DNN architecture based on a stacked autoencoder for the
  reduction of feature space dimension and a feed-forward classifier for
  multi-label classification with $\argmax$ functions to convert multi-label
  classification results into multi-class classification ones. We also describe
  the demonstration of a prototype DNN-based indoor localization system for
  floor-level location estimation using real received signal strength (RSS) data
  collected at one of the buildings on the XJTLU campus. The preliminary results
  for both building/floor classification and floor-level location estimation
  clearly show the strengths of DNN-based approaches, which can provide near
  state-of-the-art performance with less parameter tuning and higher
  scalability.
\end{abstract}


\section{Introduction}
\label{sec:introduction}
In an indoor environment where there is no line-of-sight signal from global
positioning systems (GPSs), received signal strengths (RSSs) from wireless
network infrastructure can be used for localization through
\textit{fingerprinting}: For example, a vector of a pair of a service set
identifier (SSID) and an RSS for a Wi-Fi access point (AP) measured at a
location becomes its \textit{location fingerprint}. A position of a user/device
then can be estimated by finding the closest match between its RSS measurement
and the fingerprints of known locations in a database \cite{bahl00:_radar}.

When the indoor localization is to cover a large campus or a big shopping mall
where there are lots of buildings with many floors, the scalability of
fingerprinting techniques becomes an important issue. The current
state-of-the-art Wi-Fi fingerprinting techniques assume a hierarchical approach
to the indoor localization, where the building, floor, and position (e.g., a
label or coordinates) of a location are estimated one at a time. In
\cite{moreira15:_wi_fi}, for instance, building estimation is done by the
following process: Given the AP with the strongest RSS in a measured
fingerprint, we first build a subset of fingerprints where the same AP has the
strongest RSS; then, we count the number of fingerprints associated to each
building and set the estimated building to be the most frequent one from the
counting. Similar procedures are also proposed to estimate a floor inside the
building. According to the results in \cite{moreira15:_wi_fi}, the best building
and floor hit rates achieved for the UJIIndoorLoc dataset
\cite{torres-sospedra14:_ujiin} are 100\% and 94\%, respectively.

One of the major challenges in Wi-Fi fingerprinting is how to deal with the
random fluctuation of a signal, the noise from multi-path effects, and the
device dependency in RSS measurements. Unlike traditional solutions relying on
complex filtering and time-consuming manual parameter tuning, machine learning
techniques --- especially the popular deep neural networks (DNNs) --- can
provide attractive solutions to Wi-Fi fingerprinting due to less parameter
tuning and better scalability in larger environments
\cite{felix16,zhang16:_deep,nowicki17:_low_wifi}.

In this paper, we introduce a feasibility study on the \textit{Xi'an
  Jiaotong-Liverpool University (XJTLU) Campus Information and Visitor Service
  System} as a test bed for large-scale location-aware services in access. We
report preliminary results from the investigation on the use of DNNs for
hierarchical building/floor classification and floor-level location estimation,
which we carried out as part of this feasibility study. We also describe the
demonstration of a prototype DNN-based indoor localization system for
floor-level location estimation using real RSS data, which we measured at one of
the buildings on the XJTLU campus.

The outline of the rest of the paper is as follows: In
Sec.~\ref{sec:xjtlu-camp-inform}, we introduce the feasibility study project on
the XJTLU Campus Information and Visitor Service System. In
Sec.~\ref{sec:hier-class-build-floor}, we discuss the hierarchical
classification of building and floor based on a multi-label classifier with
$\argmax$ functions. Sec.~\ref{sec:demonstr-dnn-based} describes the
demonstration of a prototype DNN-based indoor localization system for
floor-level location estimation. Sec.~\ref{sec:summary} summarizes our work
described in this paper.

\section{XJTLU Campus Information and Visitor Service System: A Test Bed for
  Large-Scale Location-Aware Services in Access}
\label{sec:xjtlu-camp-inform}
Location awareness is one of enabling technologies for future smart and green
cities; understanding where people spend their times and how they interact with
environments is critical to realizing this vision. At XJTLU, a group of
researchers from Electrical and Electronic Engineering, Computer Science \&
Software Engineering, and Urban Planning, together with an external researcher
from Electrical and Electronic Engineering of City University of London, U.K.,
has been carrying out a feasibility assessment and road mapping for
\textit{XJTLU Campus Information and Visitor Service System} with the aim of
identifying key component technologies and preparing plans for its
implementation as a test bed for large-scale location-aware services in access
and its use cases for behavioral study of students and visitors on the campus.

\begin{figure}[!tb]
  \begin{center}
    \includegraphics[angle=-90,width=.95\linewidth]{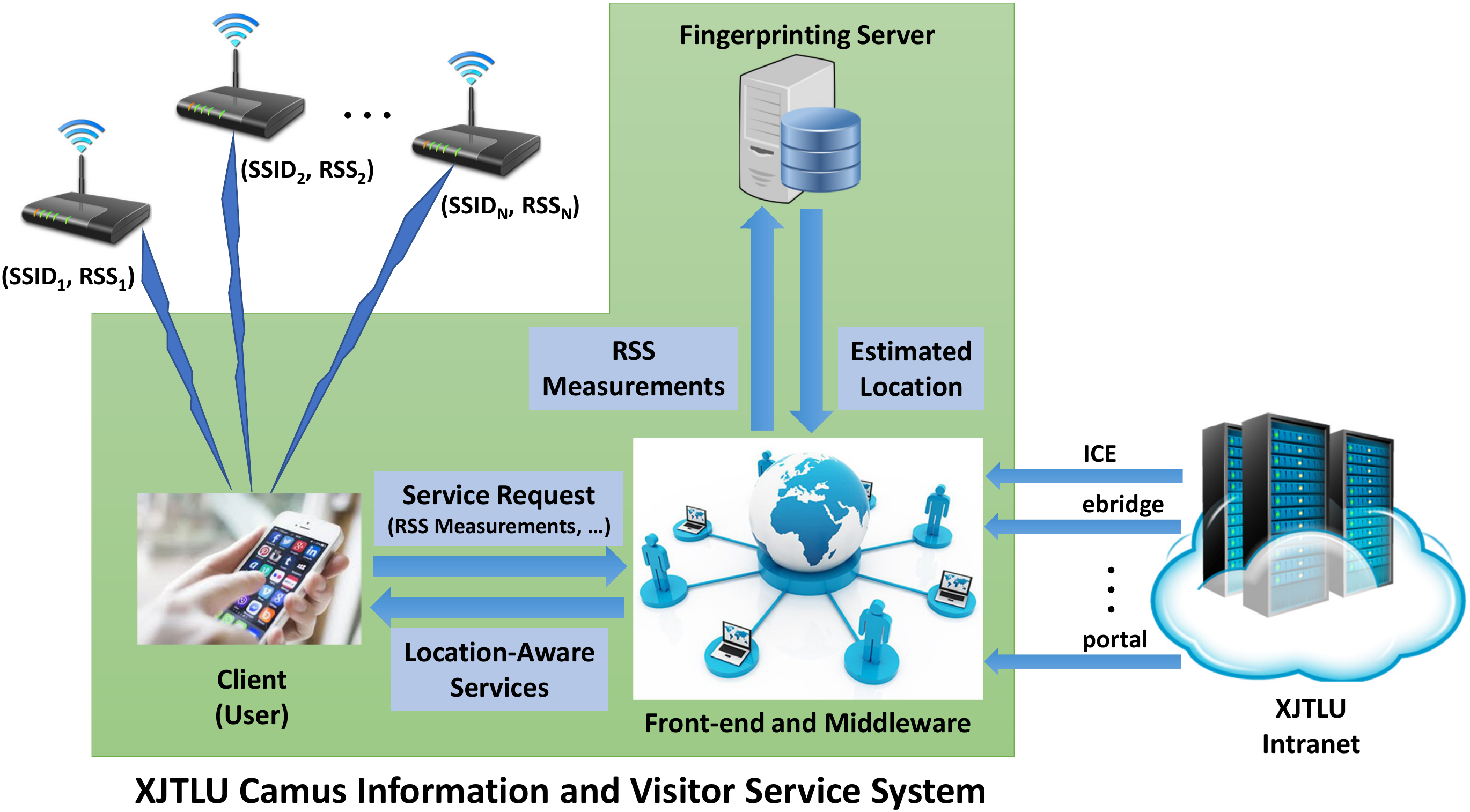}\\
    \vspace{0.2cm}
    {\scriptsize (a)}
  \end{center}
  \begin{minipage}{0.525\linewidth}
    \centering
    \includegraphics[angle=-90,width=\linewidth]{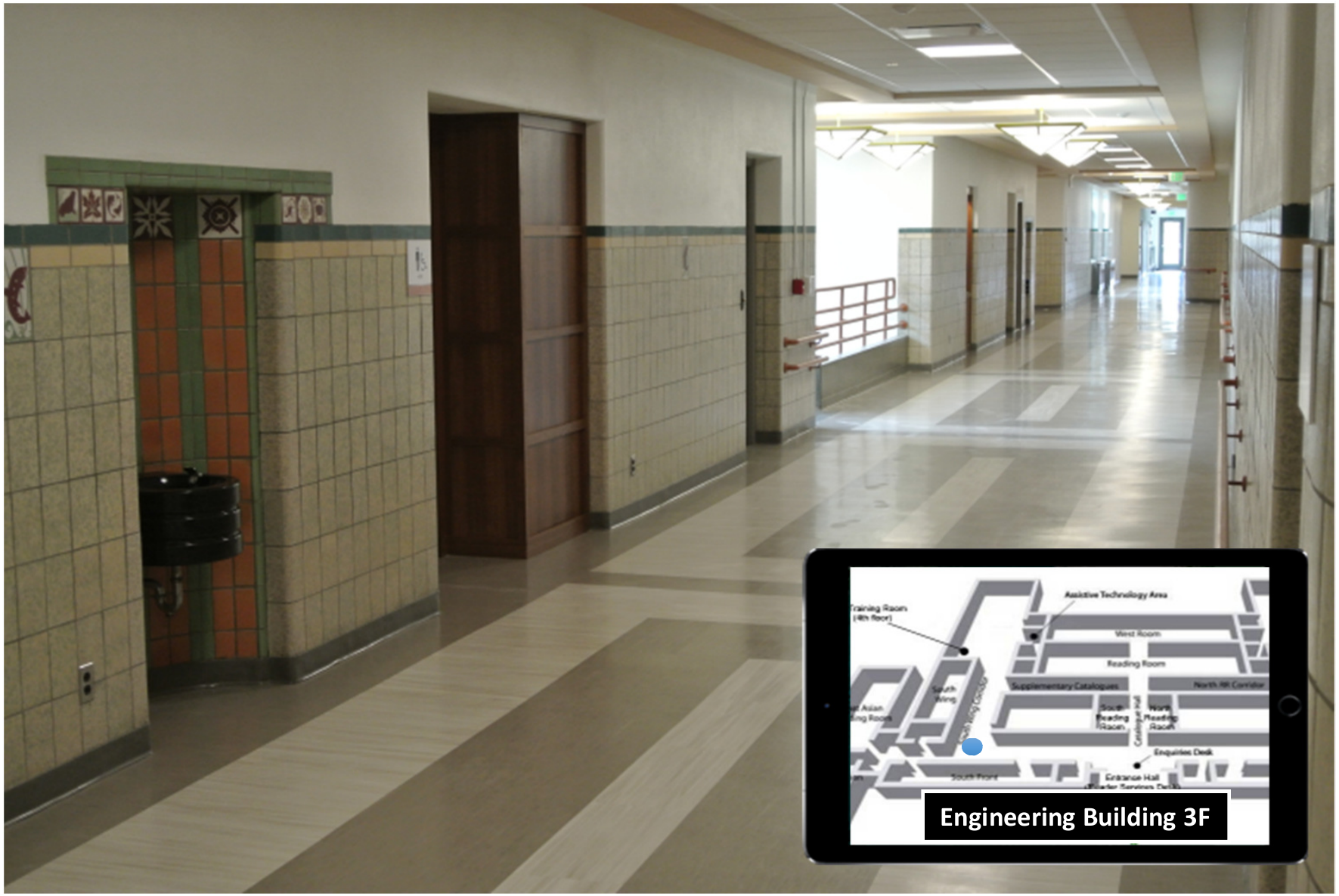}
  \end{minipage}
  \hfill
  \begin{minipage}{0.45\linewidth}
    \centering
    \includegraphics[angle=-90,width=\linewidth]{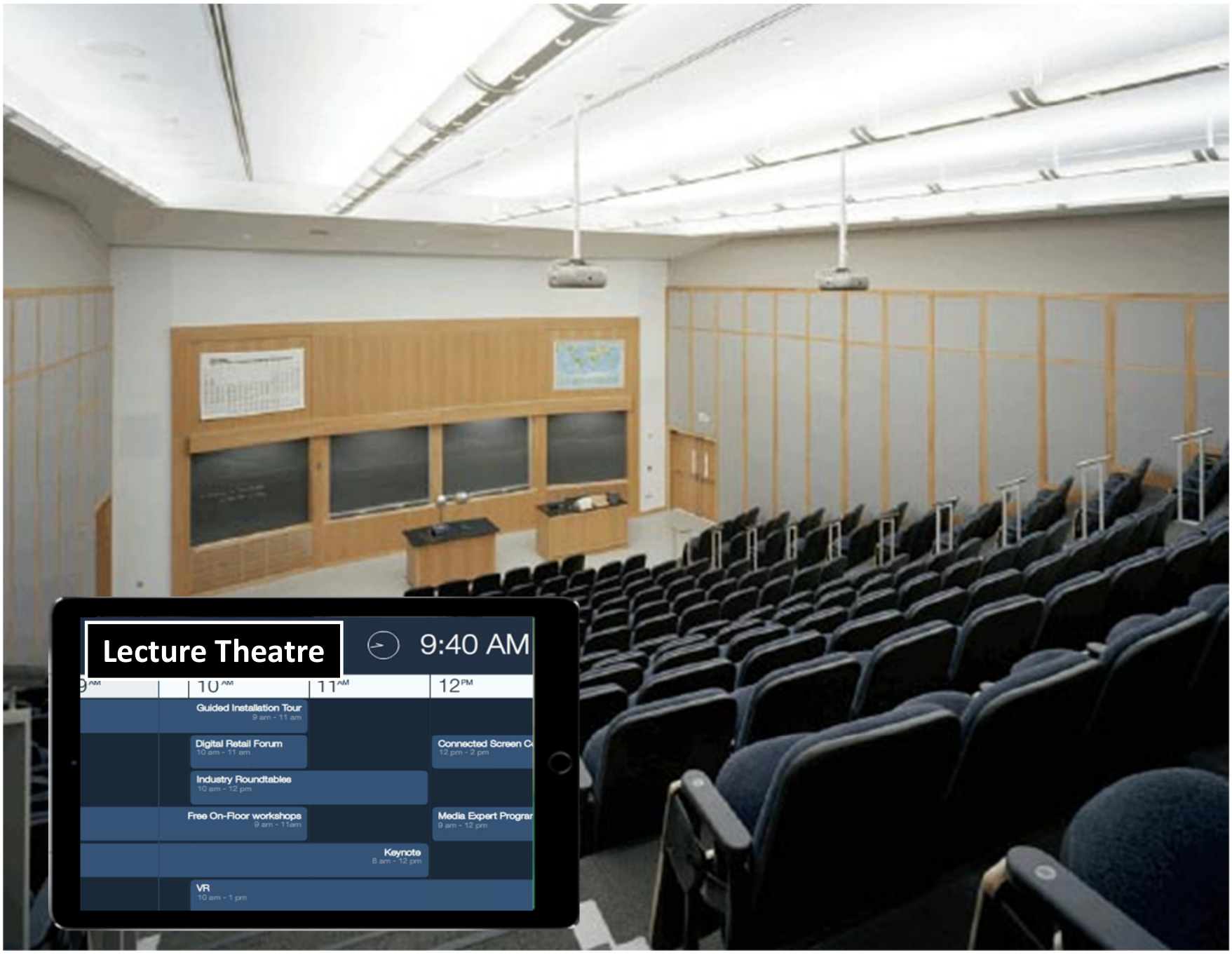}
  \end{minipage}
  \begin{center}
    {\scriptsize (b)}
  \end{center}
  \caption{ XJTLU campus information and visitor service: (a) Architecture and
    (b) location-aware service examples.}
  \label{fig:xjtlu_civs}
\end{figure}

Fig.~\ref{fig:xjtlu_civs} shows an overall architecture of the XJTLU Campus
Information and Visitor Service System and location aware service examples
provided by the system.
At the core of the system is indoor localization based on wireless
fingerprinting, which utilizes RSSs from Wi-Fi APs to estimate a user location:
As shown in Fig.~\ref{fig:xjtlu_civs}~(a), RSSs from nearby APs are submitted by
a user or an App running background in the user's mobile device to the system as
part of a service request. The submitted RSS measurements, then, are compared
with the RSS samples of known locations (i.e., location fingerprints) stored in
the database at the fingerprinting server to find the closest match; the
location ID of the closest match (e.g., ``EB306''\footnote{It means the room 306
  on the third floor of the EB building.})  is returned as the user's estimated
location. Based on the estimated location, the system provides location-aware
services by integrating existing data and services available on XJTLU Intranet
and tailoring them for the location as shown in Fig.~\ref{fig:xjtlu_civs}~(b).

\section{Hierarchical Classification of Building and Floor using A Multi-Label
  Classifier}
\label{sec:hier-class-build-floor}
\begin{figure}[!tb]
  \begin{center}
    \includegraphics[angle=-90,width=.8\linewidth]{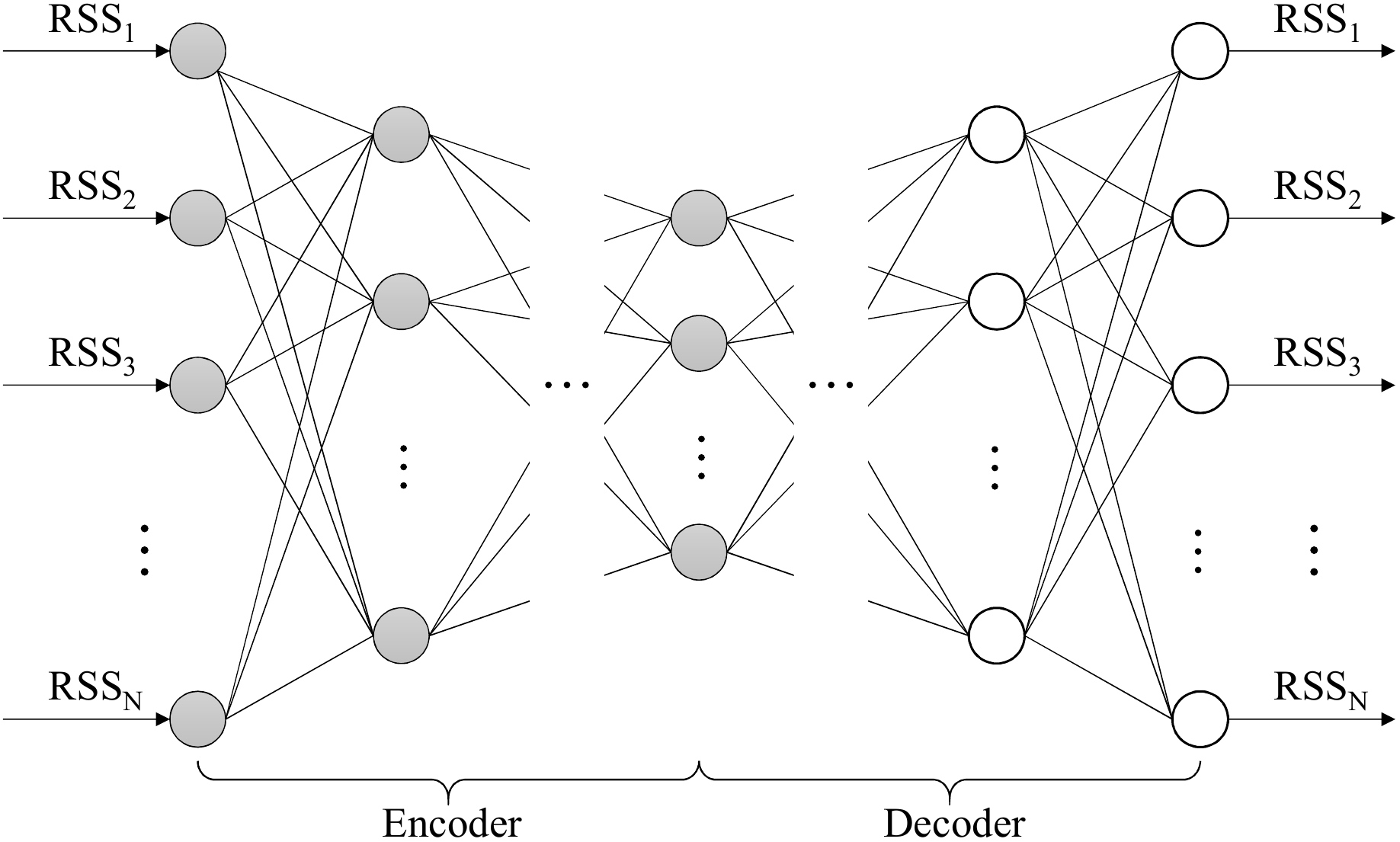}\\
    {\scriptsize (a)}\\
    \includegraphics[angle=-90,width=.9\linewidth]{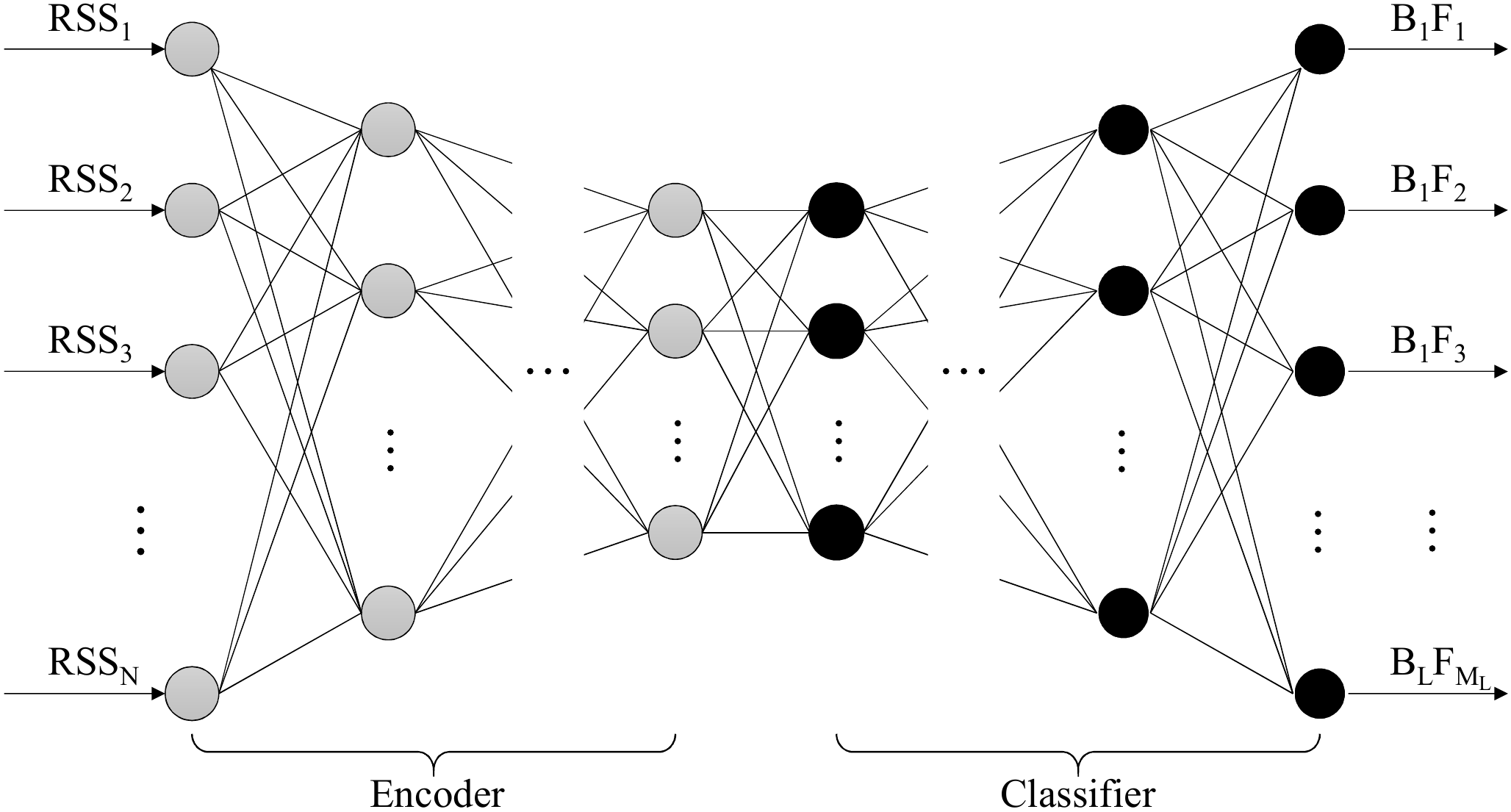}\\
    {\scriptsize (b)}
  \end{center}
  \caption{A DNN architectures for building/floor classification: (a) A stacked
    autoencoder (SAE) for the reduction of feature space dimension; (b) an
    overall architecture consisting of the encoder part of the SAE and a
    feed-forward classifier for multi-class classification with flattened
    building-floor labels \cite{nowicki17:_low_wifi}.}
  \label{fig:flat_bf_classifier}
\end{figure}

In \cite{nowicki17:_low_wifi}, the authors proposed a DNN system for
building/floor classification that uses a stacked autoencoder (SAE) before a
feed-forward classifier to reduce the dimension of the feature space for robust
and precise classification as shown in Fig.~\ref{fig:flat_bf_classifier}.
Note that, after training with RSSs as both input and output data, only the
gray-colored nodes are used as an encoder for feature space dimension reduction
as shown in Fig.~\ref{fig:flat_bf_classifier}~(b). The performance of the
proposed DNN system was verified with the UJIIndoorLoc dataset
\cite{torres-sospedra14:_ujiin} and is shown to be comparable to that of the
state-of-the-art systems (i.e., about 92\% of combined building and floor
recognitions).

The work reported in \cite{nowicki17:_low_wifi} is the first in applying a DNN
to the building/floor classification of large-scale indoor localization and, as
such, reveals a couple of areas of further improvement as follows:

First, the DNN architecture is not optimized enough. While they achieve 92\%
accuracy in building/floor classification with an SAE (i.e., only the encoder
part) containing three hidden layers of 256, 128, and 64 neurons and a
feed-forward classifier with two hidden layers of 128 neurons per each, we could
achieve a nearly equivalent performance with an SAE having just two hidden
layers of 64 and 4 neurons and a classifier without any hidden layer (i.e., the
output layer of the SAE is directly connected to the system output layer).

Second, the proposed DNN system in \cite{nowicki17:_low_wifi} does not take into
account the hierarchical nature of the classification problem at hand due to its
calculating the loss and the accuracy over flattened building-floor labels
(e.g.,
(B\textsubscript{i},\,F\textsubscript{j})$\rightarrow$``B\textsubscript{i}-F\textsubscript{j}'')\footnote{B\textsubscript{i}
  and F\textsubscript{j} denote building and floor labels, respectively.}. In
other words, the misclassification of building and that of floor have equal
loss during the training phase and results in the same accuracy during the
evaluation phase.

To take into account the hierarchical nature of the building/floor
classification, we are now considering two possible approaches: One approach is
the use of a hierarchical loss function (e.g., a loss function with different
weights for building and floor) with the existing multi-class classifier DNN
architecture and flattened labels. The other is the use of \textit{multi-label
  classification} with $\argmax$ functions to convert results back to those of
multi-class classification.\footnote{In multi-class classification (also called
  \textit{single-label classification}), an instance is associated with only a
  single label from a set of disjoint labels; in multi-label classification, on
  the other hand, an instance may be associated with multiple labels
  \cite{tsoumakas07:_multi}.}

As for the first approach, we found that the hierarchical loss function for
flattened labels does not provide a well-defined gradient function, which forces
us to use evolutionary algorithms (e.g., genetic algorithm (GA)
\cite{Goldberg:89} and particle swarm optimization (PSO)
\cite{kennedy95:_partic}) for training of DNN weights. Due to its many tradeoffs
between complexity and flexibility resulting from the use of evolutionary
algorithms in DNN weight training, this approach could be an interesting topic
for long-term research.

\begin{figure}[!tb]
  \begin{center}
    \includegraphics[angle=-90,width=\linewidth]{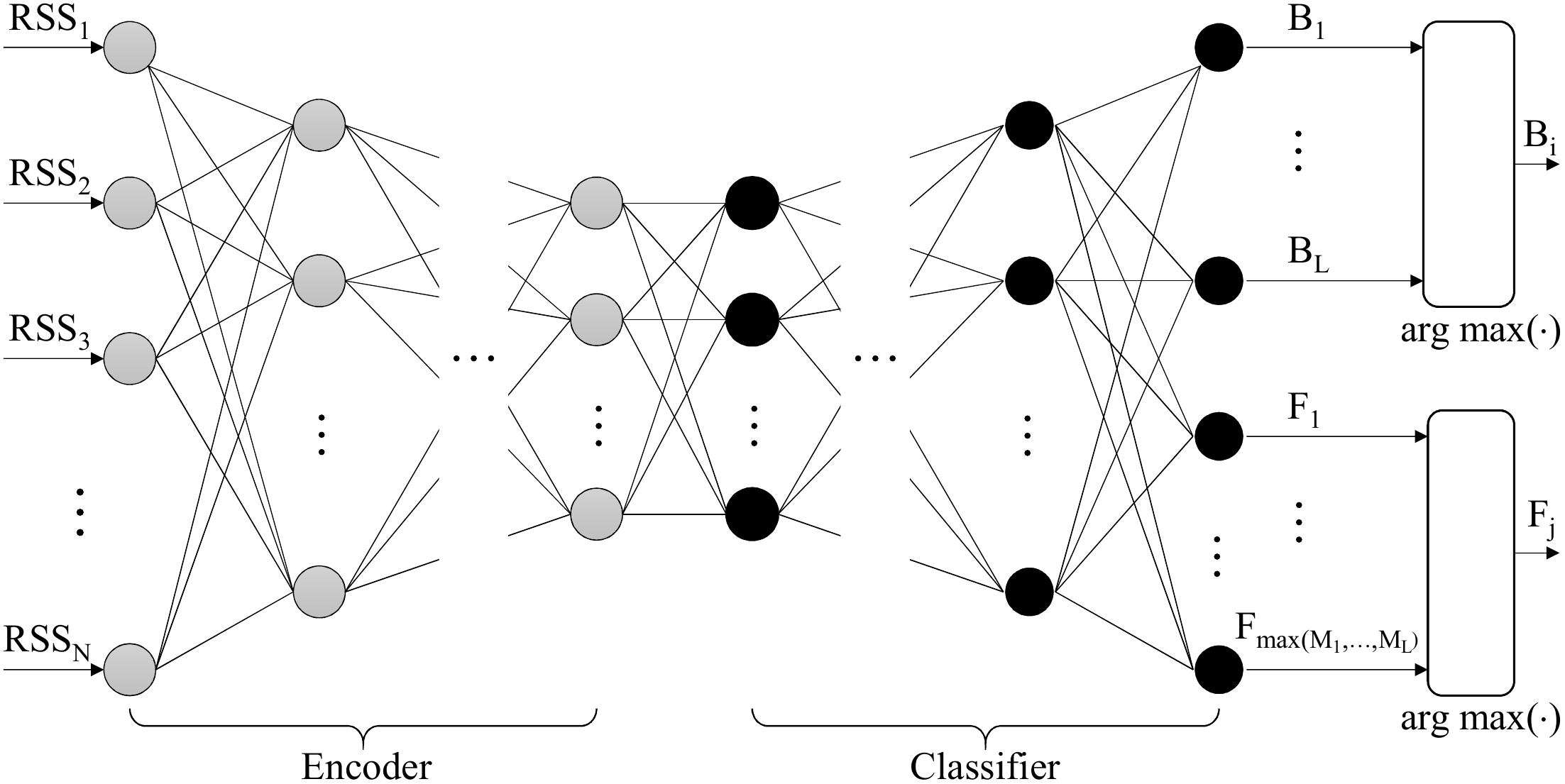}
  \end{center}
  \caption{A DNN architectures for hierarchical building/floor classification
    based on an SAE for the reduction of feature space dimension and a
    feed-forward classifier for multi-label classification.}
  \label{fig:hierarchical_bf_classifier}
\end{figure}

In case of the use of the multi-label classification, applying different weights
to building and floor labels during the training phase is rather
straightforward, and it is also easy to convert the results of multi-label
classification into those of multi-class classification using $\argmax$
functions. Fig.~\ref{fig:hierarchical_bf_classifier} shows a newly proposed
hierarchical building/floor classifier based on multi-label classification
framework with $\argmax$ functions.
%

The classification of building and floor with the proposed architecture shown in
Fig.~\ref{fig:hierarchical_bf_classifier} is achieved as follows: The three
building and five floor identifiers in the UJIIndoorLoc dataset are one-hot
encoded into an eight-dimensional vector (e.g., ``001$|$01000'' for the third
building and the second floor within it) and classified with different class
weights for buildings (i.e., for the first three digits) and for floors (i.e.,
for the last five digits); the one-hot-encoded vector from the multi-label
classifier is split into a three-dimensional building and a five-dimensional
floor vectors, and the index of a maximum value of each vector is returned as a
classified class by the $\argmax$ function.

The advantages of the proposed building/floor classifier is two-fold: First, the
number of output nodes is smaller compared to that of
\cite{nowicki17:_low_wifi}: In case of the UJIIndoorLoc dataset, the number of
output nodes is eight (i.e., the number of buildings (3) plus the maximum of the
numbers of floors for buildings (5)), while that of \cite{nowicki17:_low_wifi}
is thirteen; the difference could be much larger for large-scale access where
there are lots of buildings (i.e., $\gg 3$) with many floors.

Second, as described, the allocation of class weights can be done separately for
buildings and floors, which provides more degrees of freedom in
classification.\footnote{Note that the class weights are originally for cost
  adjustment for imbalanced data, not for hierarchical classification.} In this
regard, we are investigating the dependence of building and floor accuracies as
well as overall accuracy on class weights through extensive experiments.
Table~\ref{tbl:hierarchical_bf_classification} shows preliminary results, and
Table~\ref{tbl:hierarchical_bf_classification_dnn_parameters} summarizes common
DNN parameter values for those experiments.
\begin{table}[!tb]
  \caption{Preliminary Results of Hierarchical Building/Floor Classification}
  \label{tbl:hierarchical_bf_classification}
  \centering
  \begin{tabular}{|r|r|r|r|r|}
    \hline
    \multicolumn{2}{|c|}{Class Weight} & \multicolumn{3}{c|}{Accuracy}  \\ \hline
    \multicolumn{1}{|c|}{Building} & \multicolumn{1}{c|}{Floor} &
                                                                  \multicolumn{1}{c|}{Overall}
                                                                &
                                                                  \multicolumn{1}{c|}{Building}
                                                                &
                                                                  \multicolumn{1}{c|}{Floor} \\ \hline\hline
    1 & 1 & 8.991899e-01 & 9.918992e-01 & 9.036904e-01 \\ \hline
    2 & 1 & 9.063906e-01 & 9.918992e-01 & 9.090909e-01 \\ \hline
    5 & 1 & 9.207921e-01 & 9.972997e-01 & 9.234923e-01 \\ \hline
    \textbf{10} & \textbf{1} & \textbf{9.333933e-01} & \textbf{9.972997e-01} & \textbf{9.342934e-01} \\ \hline
    20 & 1 & 9.216922e-01 & 9.936994e-01 & 9.225923e-01 \\ \hline
  \end{tabular}
\end{table}
\begin{table}[!tb]
  \caption{DNN Parameter Values for Hierarchical Building/Floor Classification}
  \label{tbl:hierarchical_bf_classification_dnn_parameters}
  \centering
  \begin{tabular}{|l|l|}
    \hline
    \multicolumn{1}{|c|}{DNN Parameter} & \multicolumn{1}{c|}{Value} \\ \hline\hline
    Ratio of Training Data to Overall Data & 0.70 \\
    Number of Epochs & 20 \\
    Batch Size & 10 \\ \hline
    SAE Hidden Layers & 64-8-64 \\
    SAE Activation & Rectified Linear (ReLU) \\
    SAE Optimizer & ADAM \cite{kingma17:_adam} \\
    SAE Loss & Mean Squared Error (MSE) \\ \hline
    Classifier Hidden Layers & None \\
    Classifier Optimizer & ADAM \\
    Classifier Loss & Binary Crossentropy \\
    Classifier Dropout Rate & 0.20 \\ \hline
  \end{tabular}
\end{table}

According to the results shown in
Table~\ref{tbl:hierarchical_bf_classification}, the class weights of 10 and 1
for buildings and floors provide the best results, but the dependence of
accuracies on the class weights seems not to be big. This suggests that we need
to carry out a more systematic investigation on this before making any
meaningful conclusions.

\section{Demonstration of A DNN-Based Indoor Localization System for Floor-Level
  Location Estimation}
\label{sec:demonstr-dnn-based}
\begin{figure}[!tb]
  \begin{center}
	\includegraphics[width=.8\linewidth]{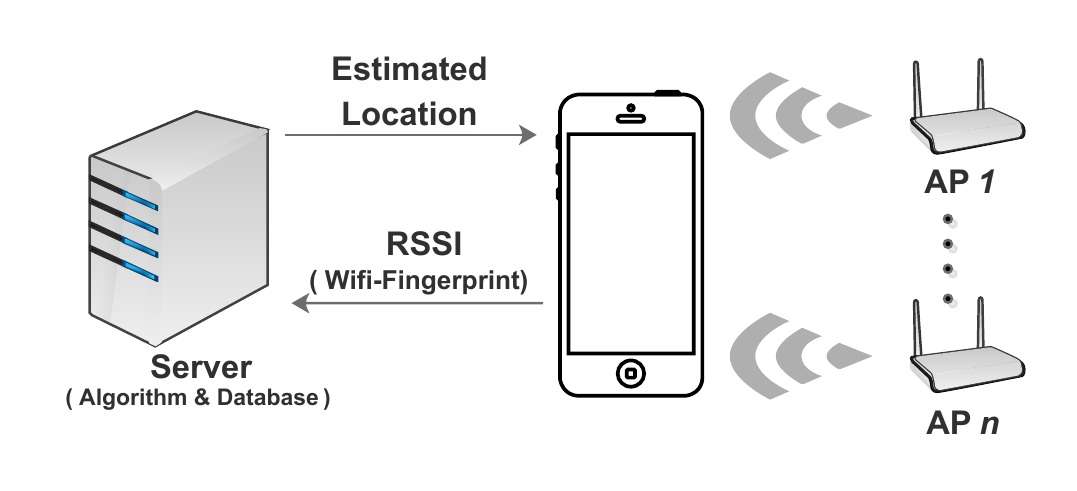}
  \end{center}
  \caption{A prototype of DNN-based indoor localization system for floor-level
    location estimation.}
  \label{fg:indoor_localization_system}
\end{figure}

To show the feasibility of DNN-based indoor localization, we have implemented a
prototype shown in Fig.~\ref{fg:indoor_localization_system} and demonstrated
floor-level location estimation with real RSS data measured on the fourth floor
of the EE building on the north campus of XJTLU.

The fingerprinting server is implemented using \textit{Flask} \cite{flask}, a
Python-based microframework for web development, with \textrm{SQLite}
\cite{sqlite} database engine; we chose Flask because we also used Python-based
deep learning frameworks, i.e., \textit{keras} \cite{keras} and TensorFlow
\cite{tensorflow}, for the implementation of DNN-based localization algorithms,
which are nearly identical to the one used for building/floor classification in
\cite{nowicki17:_low_wifi}. When a client, i.e., an Android App running on a
user's mobile phone, submits scanned RSSs, the server estimates its location
based on the implemented localization algorithm and returns the estimated
location back to the client.

\begin{figure}[!tb]
  \begin{center}
    \includegraphics[angle=-90,width=\linewidth]{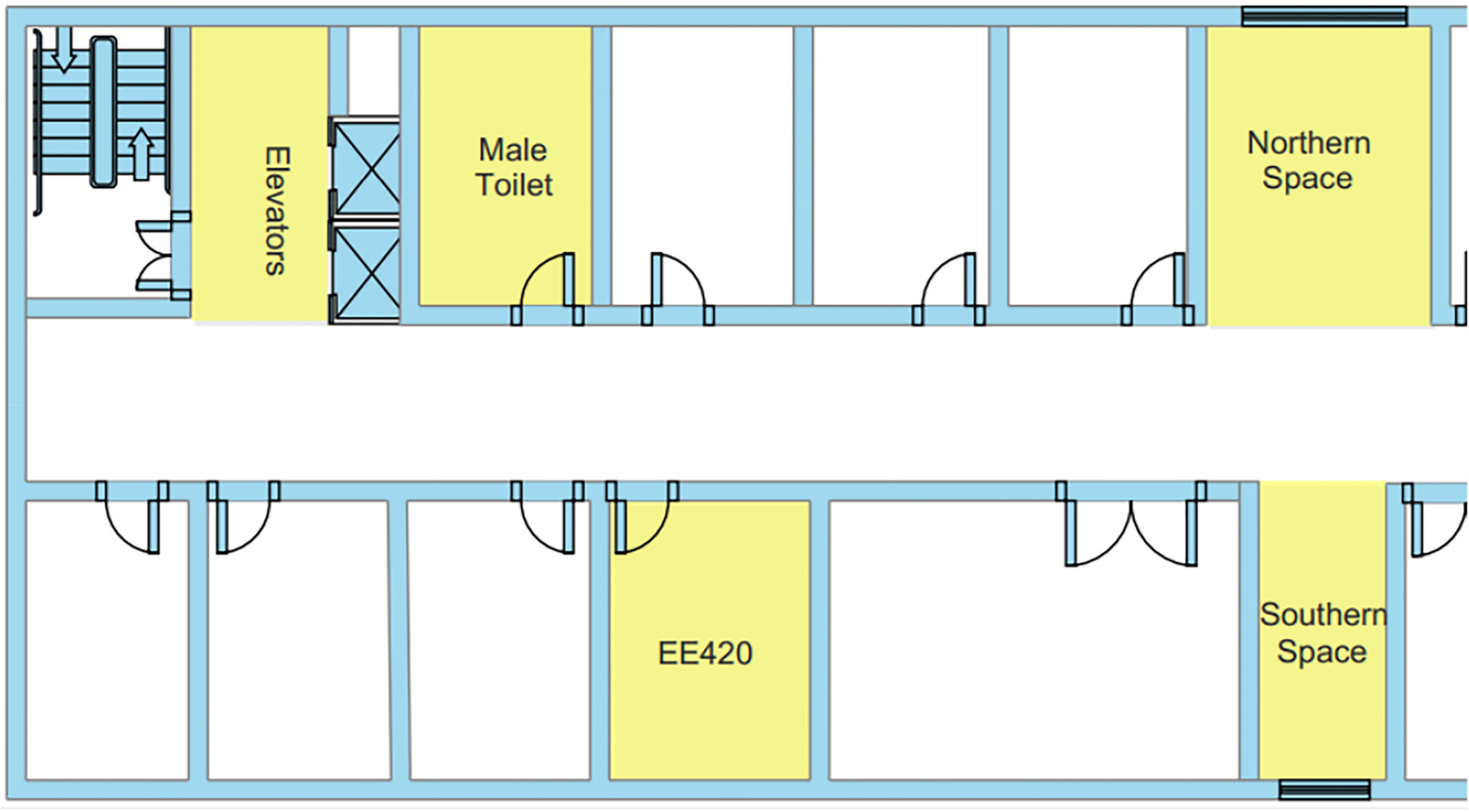}
  \end{center}
  \label{fg:floor_plan}
  \caption{A partial layout of the fourth floor of the EE building on the north
    campus of XJTLU.}
\end{figure}

Fig.~\ref{fg:floor_plan} shows a floor layout of the fourth floor of the EE
building where Wi-Fi fingerprints were collected.
The areas highlighted in yellow are part of seven locations selected to collect
data and test the prototype; both isolated rooms and open public spaces are
considered. During the measurement, we used Android mobile phones from different
brands to take into account device dependency of Wi-Fi fingerprints. From around
200 APs detected, we collected more than 4000 fingerprints at different
locations with labels.

\begin{figure}[!tb]
  \begin{center}
    \includegraphics[width=\linewidth]{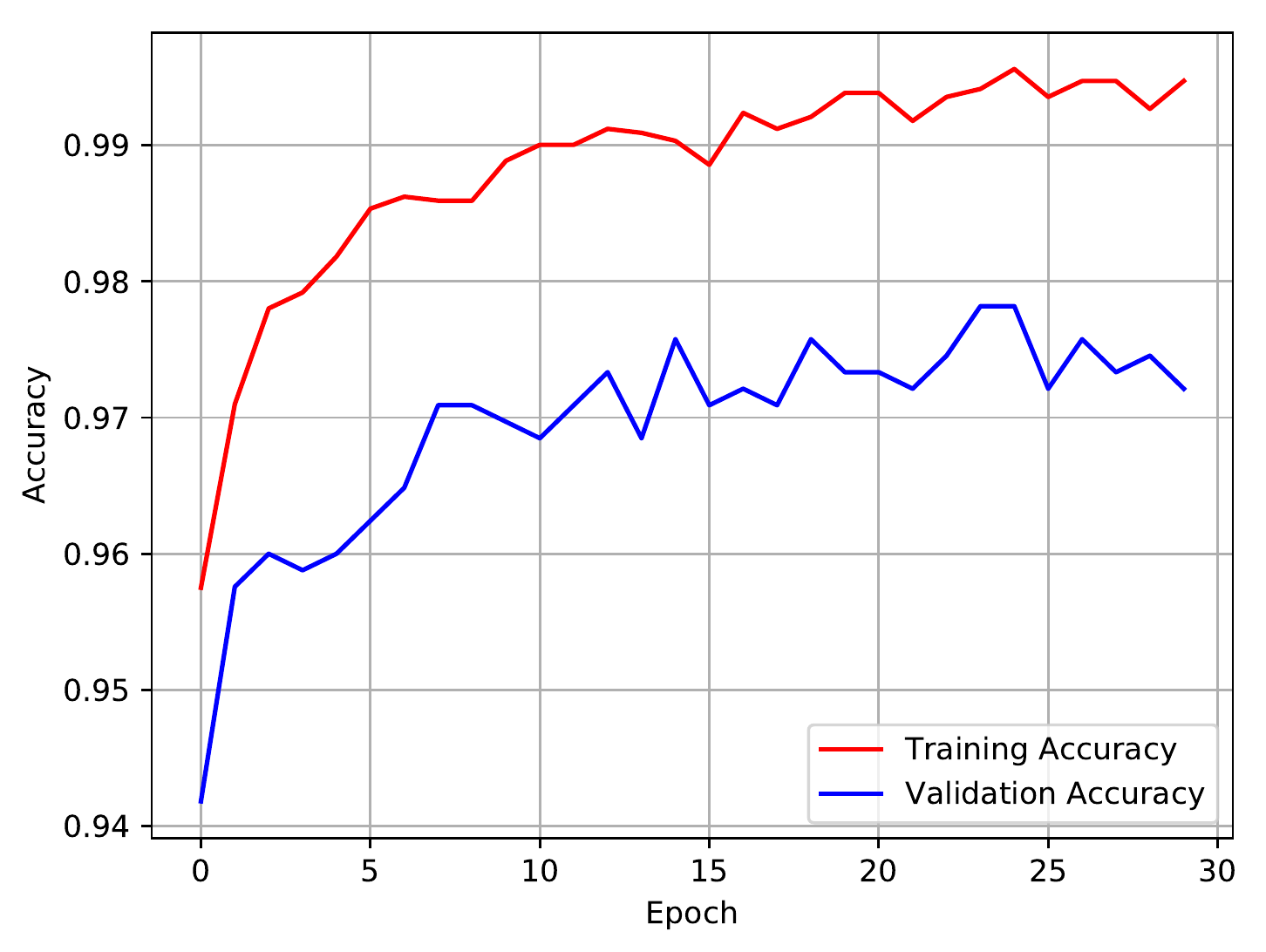}
  \end{center}
  \caption{Training and validation accuracy of floor-level location estimation
    by the DNN-based indoor localization system with real RSS data.}
  \label{fig:floor-level_localization}
\end{figure}

Fig.~\ref{fig:floor-level_localization} shows the training and validation
accuracy of floor-level location estimation by the DNN-based indoor localization
system with the collected RSS dataset. As shown in the figure, the DNN
architecture used for building/floor classification in
\cite{nowicki17:_low_wifi} can achieve more than 97\% accuracies after
thirteenth epoch in floor-level localization with both training and validation
datasets. After training and validation, we applied the system to a training
dataset of 253 fingerprints and obtained the accuracy of 0.97198, which is the
average of five trials. Note that, even though the testing accuracy is pretty
high, it may be lower if we test with a much larger dataset.
\begin{table}[!tb]
  \caption{DNN Parameter Values for Floor-Level Location Estimation}
  \label{tbl:floor-level_localization_dnn_parameters}
  \centering
  \begin{tabular}{|l|l|}
    \hline
    \multicolumn{1}{|c|}{DNN Parameter} & \multicolumn{1}{c|}{Value} \\ \hline\hline
    Ratio of Training Data to Overall Data & 0.75 \\

    Batch Size & 10 \\ \hline
    SAE Hidden Layers & 128-64-32-64-128 \\
    SAE Activation & Hyperbolic Tangent (TanH) \\
    SAE Optimizer & ADAM \\
    SAE Loss & MSE \\ \hline
    Classifier Hidden Layers & 64-32-7 \\
    Classifier Activation & ReLU \\
    Classifier Optimizer & AdaGrad \cite{duchi11:_adapt} \\
    Classifier Loss & Cross Entropy \\
    Classifier Dropout Rate & 0.50 \\
    Classifier Epochs & 30 \\ \hline
  \end{tabular}
\end{table}

\balance 

\section{Summary}
\label{sec:summary}
In this paper we have introduced the feasibility study project on the XJTLU
Campus Information and Visitor Service system, which will serve as a test bed
for large-scale location-aware services in access, and reported preliminary
results of our investigation on the use of DNNs for building/floor
classification and floor-level location estimation. The preliminary results for
both building/floor classification and floor-level location estimation clearly
show the strengths of DNN-based approaches, including immunity against signal
fluctuation, noise effects, and device dependency, no need of finding the best
match against every fingerprint in a database after DNN training, and the
elimination of time-consuming manual parameter tuning.

Still, further study is needed for hierarchical building/floor classification
and more scalable \& higher resolution floor-level localization.

Note that the implemented DNN models, results of performance evaluation, and
collected RSS datasets are available online at the project home page:
\url{http://kyeongsoo.github.io/research/projects/indoor_localization/index.html}.

\section*{Acknowledgment}
This work was supported in part by Xi'an Jiaotong-Liverpool University (XJTLU)
Research Development Fund (under Grant RDF-14-01-25), Summer Undergraduate
Research Fellowships programme (under Grant SURF-201739), Research Institute for
Smart and Green Cities Seed Grant Programme 2016-2017 (under Grant
RISGC-2017-4), and Centre for Smart Grid and Information Convergence.


\end{document}